\documentclass[onecolumn,aps]{revtex4}

\usepackage{amssymb} 
\usepackage{amsmath} 
\usepackage{amsbsy} 
\usepackage{bm} 
\usepackage{graphicx}

\begin{document} 

\title{A corrected AIC for the selection of seemingly unrelated regressions models} 
\author{J.\ L. van Velsen}
\affiliation{Dutch Ministry of Justice, Research and Documentation Centre (WODC), P.\ O.\ Box 20301, 2500 EH The Hague, The Netherlands}
\email{j.l.van.velsen@minjus.nl}
\begin{abstract}
A bias correction to Akaike's information criterion (AIC) is derived for seemingly unrelated regressions models. The correction is of particular use when the sample size is not much larger than the number of fitted parameters. A small-sample simulation study indicates that the bias-corrected AIC (AICc) provides better model choices than other model selection criteria. 
\end{abstract}

\maketitle

\section{Introduction}

The selection of a model from a set of fitted candidate models requires objective data-driven criteria. One such criterion often used in practice is Akaike's information criterion (AIC), which was designed to be an asymptotically unbiased estimator of the expected Kullback-Leibler information of a fitted model \cite{akaike73}. In finite samples, AIC has a non-vanishing bias that depends on the number of fitted parameters. This limits its effectiveness as a model selection criterion, particularly in instances where the sample size is not much larger than the number of fitted parameters of the most complex candidate model. For such instances, Hurvich and Tsai \cite{hurvich89} extended the bias-corrected AIC (AICc) originally suggested by Sugiura \cite{sugiura78} for linear regression models, to non-linear regression models and autoregressive models. Also, Hurvich and Tsai \cite{hurvich89} demonstrated the small-sample superiority of AICc over AIC as a model selection criterion. Since then, AICc has been extended to many other models, such as autoregressive moving average models \cite{hurvich90}, vector autoregressive models \cite{hurvich93} and multivariate linear regression models \cite{bedrick94}.  

The objective of this work is to define AICc for seemingly unrelated regressions models. These are models of multiple response variables that follow a joint distribution \cite{zellner62,srivastava87}. In contrast to the multivariate linear regression model of Ref.\ \cite{bedrick94}, the response variables of a seemingly unrelated regressions model do not need to depend on the same covariates. Seemingly unrelated regressions models play a central role in econometrics \cite{goldberger91} but also appear in other contexts \cite{verbyla88,rochon96,andersson01}. 
  
The remainder of this paper is organized as follows. In Sec.\ \ref{aicaicc}, the bias of AIC is calculated in seemingly unrelated regressions models with the assumption that the candidate model is either correctly specified or overspecified. The same assumption is required for AIC to be asymptotically unbiased \cite{linhart86} and has been used to calculate its bias in finite samples in other models \cite{hurvich90,hurvich93,bedrick94}. Expanded in inverse powers of the sample size $N$, the bias of AIC (${\mathcal B}_{\rm AIC}$) takes the form ${\mathcal B}_{\rm AIC}=-N^{-1}\beta(\Sigma_{0})+o(N^{-1})$, where the positive coefficient $\beta(\Sigma_{0})=O(1)$ depends on the unknown true $p \times p$ covariance matrix $\Sigma_{0}$ of the $p$ response variables. In Sec.\ \ref{aicaicc}, a lower bound $\beta^{*}>0$ of $\min_{\Omega}\beta(\Omega)$, where the minimization is over all $p \times p$ symmetric positive definite matrices $\Omega$, is found in terms of the number of fitted parameters and AICc is defined as ${\rm AICc}={\rm AIC}+N^{-1}\beta^{*}$. The performance of AICc as a model selection criterion is simulated in Sec.\ \ref{simulation} and compared to that of AIC and the Bayesian information criterion (BIC) of Schwarz \cite{schwarz78}. Finally, we give some concluding remarks in Sec.\ \ref{discussion}. Details about the calculation of $\beta(\Sigma_{0})$, its lower bound $\beta^{*}$ and the simulation study are given in, respectively, Appendices \ref{AppendixA}, \ref{AppendixB} and \ref{AppendixC}. Appendix \ref{AppendixC} also holds additional simulation results.  

\section{AIC and AIC{\tiny c}}
\label{aicaicc}

We consider the seemingly unrelated regressions model 
\begin{equation}
Y=ZB+U.
\label{model}
\end{equation}
Here, $Y$ is an $N \times p$ matrix of $p$ response variables on $N$ subjects, $Z$ is a known $N \times M$ matrix of $N$ values of $M$ covariates, each row of the $N \times p$ matrix $U$ has independent $N_{p}(0,\Sigma)$ distribution, and $B$ is an $M \times p$ matrix holding $K \le Mp$ regression coefficients and $(Mp-K)$ zeroes. The restriction $B_{ij}=0$ means that response variable $y_{j}$ of the $j$-th column of $Y$ does not depend on covariate $z_{i}$ of the $i$-th column of $Z$. The entries of the elements of the $j$-th column of $B$ that are not restricted to zero, are collected in the set $\mathcal{J}_{j}$. Each column of the matrix $B$ holds at least one regression coefficient, which means that $\mathcal{J}_{j}$ is non-empty for all $j$. Throughout this work, we assume that the $M \times M$ matrix $Z^{\rm T}Z$ is positive definite, that $p$ and $M$ do not scale with $N$, and that $\lim_{N \rightarrow \infty} N^{-1}Z^{\rm T}Z$ is finite and positive definite. 

Suppose that $Y$ is not generated by the model of Eq.\ (\ref{model}), but by the model  
\begin{equation}
Y=Z_{0}B_{0}+\mathcal{E}. 
\label{operating}
\end{equation}
Here, $Z_{0}$ is an $N \times M_{0}$ matrix of $N$ values of $M_{0}$ unknown true covariates, $B_{0}$ is an $M_{0} \times p$ matrix of unknown coefficients and each row of the $N \times p$ matrix $\mathcal{E}$ has independent $N_{p}(0,\Sigma_{0})$ distribution with unknown covariance matrix $\Sigma_{0}$. The entries of the non-vanishing elements of the $j$-th column of $B_{0}$ are collected in the set $\mathcal{J}_{0j}$. A measure of the discrepancy between the candidate (or approximating) model of Eq.\ (\ref{model}) and the data-generating model of Eq.\ (\ref{operating}) is the Kullback-Leibler information 
\begin{equation}
\Delta(B,\Sigma) = E_{0}\{-2 \mathcal{L}(B,\Sigma)\} = Np\ln 2\pi + N \ln {\rm Det}\Sigma  + {\rm Tr}(Z_{0}B_{0}-ZB)^{\rm T}(Z_{0}B_{0}-ZB)\Sigma^{-1}+ N{\rm Tr}\Sigma_{0}\Sigma^{-1},
\label{KL}
\end{equation}     
where $E_{0}$ denotes expectation under the data-generating model and $\mathcal{L}(B,\Sigma)$ is the log-likelihood function of the candidate model,
\begin{equation}
-2\mathcal{L}(B,\Sigma)=Np\ln 2\pi + N\ln {\rm Det}\Sigma +{\rm Tr}(Y-ZB)^{\rm T}(Y-ZB)\Sigma^{-1}.
\end{equation}
AIC is an estimator of the expected Kullback-Leibler information $E_{0}\{\Delta(\hat{B},\hat{\Sigma})\}$, where $\hat{B}$ and $\hat{\Sigma}$ are the maximum likelihood estimators of, respectively, $B$ and $\Sigma$. It is defined as the sum of $-2\mathcal{L}(\hat{B},\hat{\Sigma})$ and twice the number of fitted parameters, 
\begin{equation}
{\rm AIC}(\hat{\Sigma})=N\ln {\rm Det}\hat{\Sigma}+Np(\ln 2\pi + 1)+2K+p(p+1).
\label{aic}
\end{equation}

In Appendix \ref{AppendixA}, with the assumption that the candidate model is either correctly specified or overspecified ($Z_{0}=Z$ and $\mathcal{J}_{0i} \subseteq \mathcal{J}_{i}$ for all $i$), we demonstrate that  
\begin{equation}
\mathcal{B}_{\rm AIC} = E_{0}\{{\rm AIC}(\hat{\Sigma})\}-E_{0}\{\Delta(\hat{B},\hat{\Sigma})\} = -N^{-1}\beta(\Sigma_{0})+o(N^{-1}),
\label{BAIC}
\end{equation}
where $\beta(\Sigma_{0})=O(1)$ takes the form
\begin{equation}
\beta(\Sigma_{0})=  6K(p+1)+2{\rm Tr}({\rm Tr}_{\rm S}P_{0})^{2}-3{\rm Tr}P_{0}^{\vphantom{{\rm T}_{\rm S}}}P_{0}^{{\rm T}_{\rm S}}-3{\rm Tr}({\rm Tr}_{\rm R}P_{0}^{\vphantom{\rm T}})({\rm Tr}_{\rm R}P_{0}^{\rm T})+p(p+1)^{2}.
\label{bias}
\end{equation}
Here, the $Np \times Np$ oblique projection matrix $P_{0}$ is given by
\begin{equation}
P_{0}=X\{X^{\rm T}(\Sigma_{0}^{-1} \otimes \openone_{N})X\}^{-1}X^{\rm T}(\Sigma_{0}^{-1} \otimes \openone_{N}),
\label{P0}
\end{equation} 
where $X$ is an $Np \times K$ block-diagonal matrix of $p$ blocks of $N \times |\mathcal{J}_{i}|$ matrices $X_{i}$ holding the $|\mathcal{J}_{i}|$ columns of $Z$ corresponding to $z_{j}$ with $j \in \mathcal{J}_{i}$, 
\begin{equation}
X=\left( \begin{array}{ccc} X_{1} & 0 & 0 \\ 0 & \ddots & 0 \\ 0 & 0 & X_{p} \end{array} \right).
\end{equation}
In Eq.\ (\ref{bias}), the operators `${\rm Tr}_{\rm S}$' and `${\rm Tr}_{\rm R}$' denote partial traces over, respectively, the $N$ subjects and the $p$ response variables. Given an $Np \times Np$ matrix $A$, ${\rm Tr}_{\rm S}A$ is the $p \times p$ matrix defined componentwisely as $({\rm Tr}_{\rm S}A)_{ij}=\sum_{n=1}^{N} A_{in,jn}$, where $A_{in,jm}$ is multi-index notation for $A_{(i-1)N+n,(j-1)N+m}$. Similarly, ${\rm Tr}_{\rm R}A$ is the $N \times N$ matrix with elements $({\rm Tr}_{\rm R}A)_{nm}=\sum_{i=1}^{p} A_{in,im}$. Finally, in Eq.\ (\ref{bias}), `${\rm T}_{\rm S}$' denotes the partial transpose of subjects: $(A^{{\rm T}_{\rm S}})_{in,jm}=A_{im,jn}$. 

If $\mathcal{J}_{i}=\mathcal{J}_{j}$ for all $i$ and $j$, $\beta(\Sigma_{0})$ collapses to
\begin{equation}
\beta^{*}=3K(p+1)+2K^{2}p^{-1}+p(p+1)^{2},
\end{equation}
which equals the coefficient of the first term of the expansion of $-\mathcal{B}_{\rm AIC}$ of Ref.\ \cite{bedrick94} in inverse powers of $N$. In Appendix \ref{AppendixB}, we demonstrate that
\begin{equation}
\beta^{*} \le \min_{\Omega} \beta(\Omega),
\label{ineq}
\end{equation}
where the minimization is over all $p \times p$ symmetric positive definite matrices $\Omega$ and the equality sign is attained if and only if $\mathcal{J}_{i}=\mathcal{J}_{j}$ for all $i$ and $j$. We define AICc as 
\begin{equation}
{\rm AICc}(\hat{\Sigma})={\rm AIC}(\hat{\Sigma})+N^{-1}\beta^{*}.
\label{aicc}
\end{equation}
Because $0 < \beta^{*} \le \min_{\Omega} \beta(\Omega)$,  
\begin{equation}
\mathcal{B}_{\rm AICc}=E_{0}\{{\rm AICc}(\hat{\Sigma})\}-E_{0}\{\Delta(\hat{B},\hat{\Sigma})\}=-N^{-1}\left\{ \beta(\Sigma_{0})-\beta^{*} \right\}+o(N^{-1})
\end{equation}
satisfies
\begin{equation}
\lim_{N \rightarrow \infty} N\mathcal{B}_{\rm AIC} < \lim_{N \rightarrow \infty} N\mathcal{B}_{\rm AICc} \le 0.
\label{biascomp}
\end{equation}

\section{A simulation study}
\label{simulation}

We compare the performance of AIC, AICc and BIC in the selection of seemingly unrelated regressions models. For this purpose, 1000 samples of sizes $N=15$, $N=20$ and $N=50$ are created from the data-generating model (\ref{operating}) with $p=2$. For each sample and each criterion, the fitted candidate model with the smallest value of the criterion is selected from a set of candidate models. The matrix $Z$ holds the values of 10 covariates and its $10N$ elements are fixed after drawing them independently from $N(0,1)$. We consider 25 candidate models specified by $\mathcal{J}_{1}=\{1,\ldots,i\}$ and $\mathcal{J}_{2}=\{6,\ldots,5+j\}$, where $i$ and $j$ are integers ranging from 1 to 5. For the data-generating model, we set $Z_{0}=Z$ and take $\mathcal{J}_{01}=\{1,2\}$ and $\mathcal{J}_{02}=\{6,7\}$. The 4 non-vanishing elements of $B_{0}$ equal unity and the covariance matrix $\Sigma_{0}$ has parametrization $\Sigma_{0}=(1-\rho)\openone_{p}+ \rho j_{p}$, where $j_{p}$ is the $p \times p$ matrix of ones and $|\rho|<1$. The samples are constructed based on 1000 independent drawings of $\mathcal{E}$, where each row of $\mathcal{E}$ is independently drawn from $N_{p}(0,\Sigma_{0})$.

The candidate models are fitted with the constrained maximization (CM) algorithm \cite{oberhofer74,meng93}:
\begin{equation}
\hat{\Sigma}_{n+1}=N^{-1}(Y-Z\hat{B}_{n})^{\rm T}(Y-Z\hat{B}_{n}), \quad \mbox{where} \quad {\rm vec}(Z\hat{B}_{n})=X\{X^{\rm T}(\hat{\Sigma}_{n}^{-1} \otimes \openone_{N})X\}^{-1}X^{\rm T}(\hat{\Sigma}_{n}^{-1} \otimes \openone_{N}){\rm vec}(Y).
\label{CM}
\end{equation}
Here, $\hat{\Sigma}_{n+1}$ and $\hat{B}_{n}$ are estimators of, respectively, $\Sigma$ and $B$, $n$ is a positive integer and `${\rm vec}$' is the column-wise vectorization operator. The algorithm is started with $\hat{\Sigma}_{1}=\openone_{p}$ and terminated if $|{\rm Det}\hat{\Sigma}_{n+1}-{\rm Det}\hat{\Sigma}_{n}| \le \delta {\rm Det}\hat{\Sigma}_{n}$, with $\delta=1 \cdot 10^{-7}$. If the log-likelihood function $\mathcal{L}(B,\Sigma)$ is globally concave, then $\hat{\Sigma}_{n+1}$ and $\hat{B}_{n}$ converge to, respectively, $\hat{\Sigma}$ and $\hat{B}$ and the numerical error of $\ln{\rm Det}\hat{\Sigma}$ is of the order of magnitude of $\delta$. If $\mathcal{L}(B,\Sigma)$ is multi-modal, the CM algorithm does not necessarily converge to the global maximum, but may end up in a local maximum or a saddle point \cite{drton04,drton06}. Although multi-modality is rare, we choose several other initial estimators $\hat{\Sigma}_{1}$ and calculate $\ln{\rm Det}\hat{\Sigma}$ with a numerical error of about $10\delta$ (see Appendix \ref{AppendixC} for details). This means that the difference between two values of a criterion has a numerical error of $20N\delta$. 

\begin{figure}[h!]
\includegraphics[width=10cm]{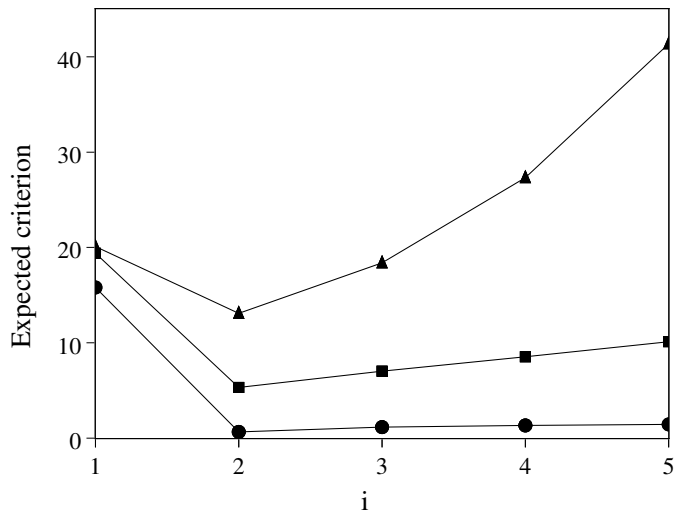}
\caption{Expected Kullback-Leibler information (triangles), AICc (squares) and AIC (circles) as a function of $i$ with $j=2$ for $N=15$ and $\rho=0.5$. The expected criteria are estimated with the same 1000 samples as the ones of Table \ref{modelselection1}. The standard error of the expected AIC (and AICc) is about $0.3$ for all $i$ and that of the expected Kullback-Leibler information ranges from $0.3$ ($i=1$) to $1.8$ ($i=5$).} 
\label{biasplot}
\end{figure}

The frequencies of selecting the 25 candidate models with the three criteria are given in Table \ref{modelselection1} for $\rho=0.5$ and $N=15$. The correct model ($i=j=2$) is more often selected with AICc than with AIC and BIC. To see how the improvement of AICc on AIC is related to the bias correction, we have plotted $E_{0}\{\Delta(\hat{\Sigma},\hat{B})\}$, $E_{0}\{{\rm AICc}(\hat{\Sigma})\}$ and $E_{0}\{{\rm AIC}(\hat{\Sigma})\}$ as a function of $i$ (with $j=2$) in Fig.\ \ref{biasplot}. The expected Kullback-Leibler information has a minimum at $i=2$ and increases rapidly with $i$ for $i>2$. This increase is more precisely followed by $E_{0}\{{\rm AICc}(\hat{\Sigma})\}$ than by $E_{0}\{{\rm AIC}(\hat{\Sigma})\}$, which explains why AIC more often selects models that are too complex. In Appendix \ref{AppendixC}, the frequencies of selecting the correct model with the three criteria are given for $\rho=0.2,0.5,0.8$ and $N=15,20,50$. The frequencies do not depend much on $\rho$ and, as expected, the improvement of AICc on AIC decreases as $N$ increases. For $N=20$, AICc and BIC perform equally well, while for $N=50$, the asymptotically consistent BIC outperforms AICc. In Appendix \ref{AppendixC}, we also demonstrate that $\delta$ is sufficiently small and that the results of Table \ref{modelselection1} are not affected by numerical errors.

\begin{table}[h!]
\caption{Frequencies of selecting the 25 candidate models with AIC, AICc and BIC in 1000 samples of size $N=15$ for $\rho=0.5$.}  
\begin{tabular}{ccccccccccccccccccc}  \hline 
   & & \multicolumn{5}{l}{AIC}               &   & \multicolumn{5}{l}{AICc}            &  & \multicolumn{5}{l}{BIC} \\
i  & & j    &        &         &       &      &   & j      &        &      &     &     &  &  j      &        &      &     &      \\ 
\cline{3-7} \cline{9-13} \cline{15-19} 
   & & 1    & 2      & 3       & 4     & 5    &   & 1      &  2     & 3    & 4   & 5   &  &  1      &  2     & 3    & 4   & 5    \\ \hline 
1  & &  0   & 1      & 0       & 0     & 0    &   & 0      &  2     & 0    & 1   & 0   &  &  0      &  2     & 0    & 0   & 0 \\ 
2  & &  3   & 241    & 75      & 60    & 66   &   & 9      &  488   & 83   & 50  & 40  &  &  7      &  385   & 78   & 53  & 58 \\  
3  & &  1   & 53     & 20      & 16    & 34   &   & 5      &  64    & 14   & 14  & 12  &  &  4      &  59    & 15   & 18  & 23 \\ 
4  & &  1   & 49     & 25      & 36    & 52   &   & 3      &  49    & 12   & 18  & 21  &  &  4      &  47    & 19   & 26  & 30 \\ 
5  & &  4   & 63     & 31      & 46    & 123  &   & 4      &  37    & 12   & 17  & 45  &  &  1      &  50    & 20   & 29  & 72 \\  \hline
\end{tabular}
\label{modelselection1}
\end{table} 

\newpage

\section{Discussion}
\label{discussion}

In the simulation study of Sec.\ \ref{simulation}, the data-generating model is finite dimensional and one of the candidate models is correctly specified. The case of an infinite dimensional data-generating model is not considered here. Although in this case the assumption of correct specification or overspecification does not hold for any candidate model, Hurvich and Tsai \cite{hurvich91} demonstrated that for linear regression models in small samples, AICc is much less biased than AIC for most choices of the data-generating model. A similar study can be done for seemingly unrelated regressions models. Also, for an infinite dimensional data-generating model, AIC and AICc are asymptotically efficient \cite{shibata80,shibata81} and, based on the results of Ref.\ \cite{hurvich91}, it can be surmised that in small samples, AICc is more efficient than AIC and BIC for most choices of the data-generating model.  

\appendix

\section{Bias of AIC}
\label{AppendixA}

In this Appendix, we demonstrate that $\mathcal{B}_{\rm AIC}=-N^{-1}\beta(\Sigma_{0}) + o(N^{-1})$, where $\beta(\Sigma_{0})=O(1)$ is given by Eq.\ (\ref{bias}). First, we calculate $\hat{\gamma}$ in the expansion
\begin{equation}
{\rm AIC}(\hat{\Sigma})-\Delta(\hat{B},\hat{\Sigma}) = -\hat{\gamma} + o_{p}(N^{-1}).
\label{decomp}
\end{equation}
Taking the expectation under the data-generating model of both sides of Eq.\ (\ref{decomp}) yields $\mathcal{B}_{\rm AIC}=-E_{0}(\hat{\gamma})+o(N^{-1})$. Second, we calculate $E_{0}(\hat{\gamma})$ and find $\beta(\Sigma_{0})$ from $E_{0}(\hat{\gamma})=N^{-1}\beta(\Sigma_{0})+o(N^{-1})$.  

\subsection{The first term of the expansion of Eq.\ (\ref{decomp})} 
\label{A1}

We consider the expansion  
\begin{equation}
\lim_{n \rightarrow \infty} \left\{{\rm AIC}(\hat{\Sigma}_{n+1})-\Delta(\hat{B}_{n},\hat{\Sigma}_{n+1})\right\} = -\hat{\eta} + o_{p}(N^{-1}),
\label{stochbias}
\end{equation}
where the estimators $\hat{\Sigma}_{n+1}$ and $\hat{B}_{n}$ of, respectively, $\Sigma$ and $B$ at the $n$-th step of the 
constrained maximization (CM) algorithm, are given by Eq.\ (\ref{CM}). Depending on the initial estimator $\hat{\Sigma}_{1}$, Drton and Richardson \cite{drton04} demonstrated that the CM algorithm may end up in a local maximum or a saddle point of $\mathcal{L}(B,\Sigma)$, rather than in the global maximum $\mathcal{L}(\hat{B},\hat{\Sigma})$. It turns out, however, that $\hat{\eta}$ does not depend on $\hat{\Sigma}_{1}$, which implies $\hat{\gamma}=\hat{\eta}$. 

Because the candidate model is either correctly specified or overspecified, the left-hand side of Eq.\ (\ref{stochbias}) can be written as
\begin{multline}
\lim_{n \rightarrow \infty} \left\{{\rm AIC}(\hat{\Sigma}_{n+1})-\Delta(\hat{B}_{n},\hat{\Sigma}_{n+1})\right\} = 2K+p(p+1) \\
- N \lim_{n \rightarrow \infty} \left\{{\rm Tr} \left(\underbrace{\Sigma_{0}-N^{-1}{\rm Tr}_{\rm S} \epsilon\epsilon^{\rm T}}_{O_{p}(N^{-1/2})}+\underbrace{N^{-1}{\rm Tr}_{\rm S} \epsilon\epsilon^{\rm T}\hat{P}_{n}^{\rm T}}_{O_{p}(N^{-1})}+\underbrace{N^{-1}{\rm Tr}_{\rm S} \hat{P}_{n}\epsilon\epsilon^{\rm T}}_{O_{p}(N^{-1})}\right) \underbrace{\hat{\Sigma}_{n+1}^{-1}}_{O_{p}(1)} \right\}, 
\label{stochbiasexpand}
\end{multline}
where $\epsilon={\rm vec}(\mathcal{E})$, `${\rm vec}$' is the column-wise vectorization operator,   
\begin{equation}
\hat{\Sigma}_{n+1}=N^{-1} {\rm Tr}_{\rm S} (\openone_{Np}-\hat{P}_{n})\epsilon\epsilon^{\rm T}(\openone_{Np}-\hat{P}_{n})^{\rm T} \quad \mbox{and} \quad
\hat{P}_{n}=X\{X^{\rm T}(\hat{\Sigma}_{n}^{-1} \otimes \openone_{N})X\}^{-1}X^{\rm T}(\hat{\Sigma}_{n}^{-1} \otimes \openone_{N}).
\label{psig}
\end{equation}
(By writing it as $N{\rm Tr}\hat{\Sigma}_{n+1}^{\vphantom{-1}}\hat{\Sigma}_{n+1}^{-1}$, the part $Np$ of ${\rm AIC}$ is absorbed in the second line of Eq.\ (\ref{stochbiasexpand}).)
The order symbols below the horizontal curly braces in Eq.\ (\ref{stochbiasexpand}) refer to the elements of the corresponding matrices. From now on, when an order symbol refers to a matrix, all of its elements are of the indicated order. (The $Np \times Np$ matrix $\hat{P}_{n}$ is $O_{p}(N^{-1})$ because $N^{-1}Z^{\rm T}Z=O(1)$.)  

The matrix $\hat{\Sigma}_{n}^{-1}$ has expansion
\begin{equation}
\hat{\Sigma}_{n}^{-1}=\Sigma_{0}^{-1}\sum_{j=0}^{Q} (-1)^{j} \left\{ (\hat{\Sigma}_{n}-\Sigma_{0})\Sigma_{0}^{-1}\right\}^{j} + o_{p}(N^{-Q/2}),
\label{exp1}
\end{equation}
where $Q$ is a non-negative integer. The expansion of Eq.\ (\ref{exp1}) holds because $p=O(1)$. Similarly, because $K=O(1)$, the matrix $\{X^{\rm T}(\hat{\Sigma}_{n}^{-1} \otimes \openone_{N})X\}^{-1}$ has expansion
\begin{multline}
\{X^{\rm T}(\hat{\Sigma}_{n}^{-1}\otimes \openone_{N})X\}^{-1}= \\
\{X^{\rm T}(\Sigma^{-1}_{0}\otimes \openone_{N})X \}^{-1}\sum_{j=0}^{Q'} (-1)^{j} \left[ X^{\rm T}\{(\hat{\Sigma}_{n}^{-1}-\Sigma^{-1}_{0})\otimes \openone_{N}\}X\{X^{\rm T}(\Sigma^{-1}_{0}\otimes \openone_{N})X\}^{-1}  \right]^{j} + o_{p}(N^{-Q'/2-1}), 
\label{exp2}
\end{multline} 
where $Q'$ is a non-negative integer.
By combining Eq.\ (\ref{exp1}) with $Q=2$, Eq.\ (\ref{exp2}) with $Q'=2$ and $\lim_{n \rightarrow \infty} \hat{\Sigma}_{n}=N^{-1}{\rm Tr}_{\rm S}(\openone_{Np}-P_{0})\epsilon\epsilon^{\rm T}(\openone_{Np}-P_{0})^{\rm T}+o_{p}(N^{-1})$, where $P_{0}=O(N^{-1})$ is given by Eq.\ (\ref{P0}), we obtain
\begin{equation}
\lim_{n \rightarrow \infty} \hat{P}_{n}=P_{0}+\hat{P}_{-3/2}+\hat{P}_{-2}+o_{p}(N^{-2}),
\label{Pdecomp}
\end{equation}
where the matrices $\hat{P}_{-3/2}=O_{p}(N^{-3/2})$ and $\hat{P}_{-2}=O_{p}(N^{-2})$ are given by
\begin{equation}
\hat{P}_{-3/2}=P_{0}\{(N^{-1}{\rm Tr}_{\rm S} \epsilon\epsilon^{\rm T}-\Sigma_{0})\otimes \openone_{N}\}(P_{0}^{\rm T}-\openone_{Np})(\Sigma_{0}^{-1} \otimes \openone_{N}) 
\end{equation}
and 
\begin{multline}
\hat{P}_{-2}=N^{-1}P_{0}\{({\rm Tr}_{\rm S}\epsilon\epsilon^{\rm T}P_{0}^{\rm T}+{\rm Tr}_{\rm S}P_{0}\epsilon\epsilon^{\rm T}-{\rm Tr}_{\rm S}P_{0}^{\vphantom{\rm T}}\epsilon\epsilon^{\rm T}P_{0}^{\rm T})\otimes \openone_{N}\}(\openone_{Np}-P^{\rm T}_{0})(\Sigma_{0}^{-1}\otimes\openone_{N}) \\ -\hat{P}_{-3/2}\{(N^{-1}{\rm Tr}_{\rm S} \epsilon\epsilon^{\rm T} - \Sigma_{0})\otimes\openone_{N}\}(\Sigma_{0}^{-1} \otimes \openone_{N}) (\openone_{Np}-P_{0}).
\end{multline}
By combining Eq.\ (\ref{exp1}) with $Q=3$ and $\lim_{n \rightarrow \infty}\hat{P}_{n}=P_{0}+\hat{P}_{-3/2}+o_{p}(N^{-3/2})$, we obtain 
\begin{equation}
\lim_{n \rightarrow \infty} \hat{\Sigma}_{n+1}^{-1}= \Sigma_{0}^{-1} \sum_{j=0}^{3} (-1)^{j} \left( \left[N^{-1} {\rm Tr}_{\rm S} \{\openone_{Np}-(P_{0}+\hat{P}_{-3/2})\}\epsilon\epsilon^{\rm T}\{\openone_{Np}-(P_{0}+\hat{P}_{-3/2})\}^{\rm T}-\Sigma_{0}\right]\Sigma_{0}^{-1}\right)^{j}+o_{p}(N^{-3/2}).
\label{Sigmadecomp}
\end{equation}

Substituting the expansions of Eqs.\ (\ref{Pdecomp},\ref{Sigmadecomp}) in the right-hand side of Eq.\ (\ref{stochbiasexpand}), expressing it as the right-hand side of Eq.\ (\ref{stochbias}), and noting that $\hat{\gamma}=\hat{\eta}$ (because $\hat{\eta}$ does not depend on $\hat{\Sigma}_{1}$), yields
\begin{equation}
\hat{\gamma}=N{\rm Tr}(\Sigma_{0}-N^{-1}{\rm Tr}_{\rm S}\epsilon\epsilon^{\rm T})\Sigma_{0}^{-1} + \hat{\gamma}_{0}+\hat{\gamma}_{-1/2}+\hat{\gamma}_{-1},
\label{hatbeta}
\end{equation} 
where $\hat{\gamma}_{0}=O_{p}(1)$, $\hat{\gamma}_{-1/2}=O_{p}(N^{-1/2})$ and $\hat{\gamma}_{-1}=O_{p}(N^{-1})$ are given by
\begin{equation}
\hat{\gamma}_{0}=-2K-p(p+1)+ {\rm Tr}({\rm Tr}_{\rm S}\epsilon\epsilon^{\rm T}P_{0}^{\rm T}+{\rm Tr}_{\rm S} P_{0}\epsilon\epsilon^{\rm T})\Sigma_{0}^{-1} + N{\rm Tr}\{(\Sigma_{0}-N^{-1}{\rm Tr}_{\rm S} \epsilon\epsilon^{\rm T})\Sigma_{0}^{-1} \}^{2},
\end{equation}
\begin{equation}
\begin{array}{lll}
\hat{\gamma}_{-1/2} & = & 2{\rm Tr}(\Sigma_{0}-N^{-1}{\rm Tr}_{\rm S} \epsilon\epsilon^{\rm T})\Sigma_{0}^{-1}({\rm Tr}_{\rm S} \epsilon\epsilon^{\rm T}P^{\rm T}_{0}+{\rm Tr}_{\rm S} P_{0}\epsilon\epsilon^{\rm T})\Sigma_{0}^{-1}+N{\rm Tr}\{(\Sigma_{0}-N^{-1}{\rm Tr}_{\rm S} \epsilon\epsilon^{\rm T})\Sigma_{0}^{-1}\}^{3} \\
                   & - & {\rm Tr}(\Sigma_{0}-N^{-1}{\rm Tr}_{\rm S} \epsilon\epsilon^{\rm T})\Sigma_{0}^{-1}({\rm Tr}_{\rm S} P_{0}^{\vphantom{\rm T}}\epsilon\epsilon^{\rm T}P^{\rm T}_{0})\Sigma_{0}^{-1} 
                    +  {\rm Tr}({\rm Tr}_{\rm S} \epsilon\epsilon^{\rm T}\hat{P}^{\rm T}_{-3/2}+{\rm Tr}_{\rm S} \hat{P}_{-3/2}\epsilon\epsilon^{\rm T})\Sigma_{0}^{-1}
\end{array}
\end{equation}
and 
\begin{equation}
\begin{array}{lll}
\hat{\gamma}_{-1} &  = & N^{-1}{\rm Tr}({\rm Tr}_{\rm S} \epsilon\epsilon^{\rm T}P^{\rm T}_{0}+{\rm Tr}_{\rm S} P_{0}\epsilon\epsilon^{\rm T})\Sigma_{0}^{-1}({\rm Tr}_{\rm S} \epsilon\epsilon^{\rm T}P^{\rm T}_{0}+{\rm Tr}_{\rm S} P_{0}\epsilon\epsilon^{\rm T}-{\rm Tr}_{\rm S}P_{0}^{\vphantom{\rm T}}\epsilon\epsilon^{\rm T}P^{\rm T}_{0})\Sigma_{0}^{-1} \\
                 &  + & 2{\rm Tr}({\rm Tr}_{\rm S}\epsilon\epsilon^{\rm T}\hat{P}^{\rm T}_{-3/2}+{\rm Tr}_{\rm S} \hat{P}_{-3/2}\epsilon\epsilon^{\rm T})\Sigma_{0}^{-1}(\Sigma_{0}-N^{-1}{\rm Tr}_{\rm S} \epsilon\epsilon^{\rm T})\Sigma_{0}^{-1} \\ 
                 &  + & {\rm Tr}({\rm Tr}_{\rm S}\epsilon\epsilon^{\rm T}\hat{P}^{\rm T}_{-2}+{\rm Tr}_{\rm S} \hat{P}_{-2}\epsilon\epsilon^{\rm T})\Sigma_{0}^{-1}+ N{\rm Tr}\{(\Sigma_{0}-N^{-1}{\rm Tr}_{\rm S} \epsilon\epsilon^{\rm T})\Sigma_{0}^{-1}\}^{4} \\
                 &  - & {\rm Tr}({\rm Tr}_{\rm S}\hat{P}_{-3/2}\epsilon\epsilon^{\rm T}P^{\rm T}_{0}+{\rm Tr}_{\rm S}P_{0}\epsilon\epsilon^{\rm T}\hat{P}^{\rm T}_{-3/2})\Sigma_{0}^{-1}(\Sigma_{0}-N^{-1}{\rm Tr}_{\rm S} \epsilon\epsilon^{\rm T})\Sigma_{0}^{-1} \\
                 &  + & 3{\rm Tr}({\rm Tr}_{\rm S} \epsilon\epsilon^{\rm T}P^{\rm T}_{0}+{\rm Tr}_{\rm S} P_{0}\epsilon\epsilon^{\rm T})\Sigma_{0}^{-1}\{(\Sigma_{0}-N^{-1}{\rm Tr}_{\rm S} \epsilon\epsilon^{\rm T})\Sigma_{0}^{-1}\}^{2} \\
                 &  - & 2{\rm Tr}({\rm Tr}_{\rm S}P_{0}^{\vphantom{\rm T}}\epsilon\epsilon^{\rm T}P^{\rm T}_{0})\Sigma_{0}^{-1}\{(\Sigma_{0}-N^{-1}{\rm Tr}_{\rm S} \epsilon\epsilon^{\rm T})\Sigma_{0}^{-1}\}^{2}. \\
\end{array}
\end{equation}

\subsection{Expectation under the data-generating model}
\label{A2}

The elements of the $Np$-dimensional Gaussian columnvector $\epsilon={\rm vec}(\mathcal{E})$ have vanishing mean and two-point average
\begin{equation}
E_{0}(\epsilon_{in}\epsilon_{jm})=(\Sigma_{0})_{ij}\delta_{nm},
\end{equation}
where $\epsilon_{in}$ is multi-index notation for $\epsilon_{N(i-1)+n}=\mathcal{E}_{in}$ and $\delta_{nm}$ is a Kronecker delta. Because $E_{0}\{N{\rm Tr}(\Sigma_{0}-N^{-1}{\rm Tr}_{\rm S}\epsilon\epsilon^{\rm T})\Sigma_{0}^{-1}\}=0$, $E_{0}(\hat{\gamma})$ takes the form
\begin{equation}
E_{0}(\hat{\gamma})=E_{0}(\hat{\gamma}_{0})+E_{0}(\hat{\gamma}_{-1/2})+E_{0}(\hat{\gamma}_{-1}).
\label{avegamma}
\end{equation}
Applying Wick's theorem, which states that the average of a product of $2g$ elements of $\epsilon$, where $g$ is a positive integer, equals the sum of products of all $\prod_{i=1}^{g}(2i-1)$ possible pairings of two-point averages, we obtain
\begin{equation}
E_{0}(\hat{\gamma}_{0})=0,
\label{ave0}
\end{equation}
\begin{equation}
E_{0}(\hat{\gamma}_{-1/2})=N^{-1} [ -6K(p+1)+3{\rm Tr}P_{0}^{\vphantom{{\rm T}_{\rm S}}}P_{0}^{{\rm T}_{\rm S}}+3{\rm Tr}({\rm Tr}_{\rm R}P_{0}^{\vphantom{\rm T}})({\rm Tr}_{\rm R}P_{0}^{\rm T})-\{p^{2}+3p+p(p+1)^{2}\} ]
\label{ave1}
\end{equation}
and 
\begin{equation}
E_{0}(\hat{\gamma}_{-1})= N^{-1} \{ 12K(p+1)+2{\rm Tr}({\rm Tr}_{\rm S}P_{0})^{2}-6{\rm Tr}P_{0}^{\vphantom{{\rm T}_{\rm S}}}P_{0}^{{\rm T}_{\rm S}}-6{\rm Tr}({\rm Tr}_{\rm R}P_{0}^{\vphantom{\rm T}})({\rm Tr}_{\rm R}P_{0}^{\rm T})+p^{2}+3p+2p(p+1)^{2} \}+o(N^{-1}).
\label{ave2}
\end{equation}
Substituting Eqs.\ (\ref{ave0},\ref{ave1},\ref{ave2}) in Eq.\ (\ref{avegamma}), yields
\begin{equation}
E_{0}(\hat{\gamma})=N^{-1}\beta(\Sigma_{0})+o(N^{-1}),
\end{equation}
where $\beta(\Sigma_{0})=O(1)$ is given by Eq.\ (\ref{bias}).
 
\section{Proof of Eq.\ (\ref{ineq})}
\label{AppendixB}

In this Appendix, we demonstrate
\begin{equation}
-3K(p+1)+2K^{2}p^{-1} \le \min_{\Omega} \left[\left\{ 2{\rm Tr}({\rm Tr}_{\rm S}P_{0})^{2}-3{\rm Tr}P_{0}^{\vphantom{\rm T}}P_{0}^{{\rm T}_{\rm S}}-3{\rm Tr}({\rm Tr}_{\rm R}P_{0}^{\vphantom{\rm T}})({\rm Tr}_{\rm R}P_{0}^{\rm T}) \right\}_{\Sigma_{0}=\Omega} \right],
\label{ineqapp}
\end{equation}
where the minimization is over all $p \times p$ symmetric positive definite matrices $\Omega$ and the equality sign is attained if and only if $\mathcal{J}_{i}=\mathcal{J}_{j}$ for all $i$ and $j$. By adding $6K(p+1)+p(p+1)^{2}$ on both sides of Eq.\ (\ref{ineqapp}), we obtain $\beta^{*} \le \min_{\Omega}\beta(\Omega)$ of Eq.\ (\ref{ineq}). 

Using ${\rm Tr}_{\rm S}P_{0}=\Sigma_{0}^{1/2}({\rm Tr}_{\rm S}\mathcal{A})\Sigma_{0}^{-1/2}$, where 
\begin{equation}
\mathcal{A}=(\Sigma_{0}^{-1/2} \otimes \openone_{N})P_{0}(\Sigma_{0}^{1/2} \otimes \openone_{N})=(\Sigma_{0}^{-1/2} \otimes \openone_{N})X\{X^{\rm T}(\Sigma_{0}^{-1} \otimes \openone_{N})X\}^{-1}X^{\rm T}(\Sigma_{0}^{-1/2} \otimes \openone_{N}),
\end{equation}
we find that ${\rm Tr}({\rm Tr}_{\rm S}P_{0})^{2}$ can be written as
\begin{equation}
{\rm Tr}({\rm Tr}_{\rm S}P_{0})^{2}={\rm Tr}({\rm Tr}_{\rm S}\mathcal{A})^{2}.
\end{equation}
From
\begin{equation}
\min_{\mathcal{C}} \left( {\rm Tr}\,\mathcal{C}^{2} \, \vline \, {\rm Tr}\,\mathcal{C}=K \right)=K^{2}p^{-1},
\label{bound1u}
\end{equation}
where the minimization is over all $p \times p$ symmetric matrices $\mathcal{C}$, we obtain
\begin{equation}
\min_{\Omega}\left[\left\{ {\rm Tr}({\rm Tr}_{\rm S}P_{0})^{2}\right\}_{\Sigma_{0}=\Omega}\right] = \min_{\Omega}\left[\left\{{\rm Tr}({\rm Tr}_{\rm S}\mathcal{A})^{2}\right\}_{\Sigma_{0}=\Omega}\right] \ge K^{2}p^{-1}.
\label{bound1c}
\end{equation}
The minimum of Eq.\ (\ref{bound1u}) is attained if and only if $\mathcal{C}_{ii}=Kp^{-1}$ and $\mathcal{C}_{ij}=0$ for all $i \neq j$. This corresponds to ${\rm Tr}_{\rm S}\mathcal{A}=Kp^{-1}\openone_{p}$, which can be reached if $\mathcal{J}_{i}=\mathcal{J}_{j}$ for all $i$ and $j$ or if $\Sigma_{0}=\openone_{p}$ and $|\mathcal{J}_{i}|=|\mathcal{J}_{j}|$ for all $i$ and $j$.  

Because 
\begin{equation}
P_{0}^{{\rm T}_{\rm S}}=(\Sigma_{0}^{1/2} \otimes \openone_{N})\mathcal{A}^{{\rm T}_{\rm S}}(\Sigma_{0}^{-1/2} \otimes \openone_{N}),
\end{equation}
${\rm Tr}P_{0}^{\vphantom{{\rm T}_{\rm S}}}P_{0}^{{\rm T}_{\rm S}}$ equals the inner product of $\mathcal{A}^{{\rm T}_{\rm S}}$ and $\mathcal{A}$:
\begin{equation}
{\rm Tr}P_{0}^{\vphantom{{\rm T}_{\rm S}}}P_{0}^{{\rm T}_{\rm S}}={\rm Tr}\mathcal{A}^{{\rm T}_{\rm S}}\mathcal{A}^{\rm T}.
\end{equation}
The squared length ${\rm Tr}\mathcal{A}\mathcal{A}^{\rm T}$ of $\mathcal{A}$ equals $K$ ($\mathcal{A}$ is an orthogonal projection matrix of rank $K$). 
The squared length of $\mathcal{A}^{{\rm T}_{\rm S}}$ equals that of $\mathcal{A}$ and we have
\begin{equation}
\max_{\Omega}\left\{ \left({\rm Tr}P_{0}^{\vphantom{{\rm T}_{\rm S}}}P_{0}^{{\rm T}_{\rm S}}\right)_{\Sigma_{0}=\Omega}\right\}= 
\max_{\Omega}\left\{ \left({\rm Tr}\mathcal{A}^{{\rm T}_{\rm S}}\mathcal{A}^{\rm T}\right)_{\Sigma_{0}=\Omega} \right\} \le \left\{{\rm Tr}\mathcal{A}{\mathcal{A}}^{\rm T}{\rm Tr}\mathcal{A}^{{\rm T}_{\rm S}}(\mathcal{A}^{{\rm T}_{\rm S}})^{\rm T}\right\}^{1/2}=K.
\label{bound2c}
\end{equation}
The upper bound of $K$ in Eq.\ (\ref{bound2c}) is attained if and only if $\mathcal{A}=\mathcal{A}^{{\rm T}_{\rm S}}$, which can be reached if $\mathcal{J}_{i}=\mathcal{J}_{j}$ for all $i$ and $j$ or if $\Sigma_{0}=\openone_{p}$.

Using ${\rm Tr}_{\rm R}P_{0}={\rm Tr}_{\rm R}\mathcal{A}$, we find
\begin{equation}
{\rm Tr}({\rm Tr}_{\rm R}P_{0}^{\vphantom{\rm T}})({\rm Tr}_{\rm R}P_{0}^{\rm T})={\rm Tr}({\rm Tr}_{\rm R}\mathcal{A})({\rm Tr}_{\rm R}\mathcal{A})^{\rm T} = \sum_{i=1}^{p}\sum_{j=1}^{p} {\rm Tr} a_{ii}^{\vphantom{\rm T}}a_{jj}^{\rm T},
\end{equation}
where $a_{ij}$ is the $ij$-th $N \times N$ submatrix of $\mathcal{A}$. The sum of the squared lengths of the $a_{ii}$'s is bounded by 
\begin{equation}
\sum_{i=1}^{p}{\rm Tr}a_{ii}^{\vphantom{\rm T}}a_{ii}^{\rm T} \le \sum_{i=1}^{p}\sum_{j=1}^{p} {\rm Tr}a_{ij}^{\vphantom{\rm T}}a_{ij}^{\rm T} = {\rm Tr}\mathcal{A}\mathcal{A}^{\rm T}=K
\end{equation} 
The upper bound $pK$ of 
\begin{equation}
\max_{\{c_{ii}\}} \left( \sum_{i=1}^{p}\sum_{j=1}^{p} {\rm Tr} c_{ii}^{\vphantom{\rm T}}c_{jj}^{\rm T} \vline \sum_{i=1}^{p}{\rm Tr}c_{ii}^{\vphantom{\rm T}}c_{ii}^{\rm T} \le K \right) \le pK,
\label{maxa}
\end{equation}
where the maximization is over $p$ symmetric $N \times N$ matrices $c_{ii}$, is attained if and only if $c_{ii}=c_{jj}$ and $\sum_{i=1}^{p} {\rm Tr}c_{ii}^{\vphantom{\rm T}}c_{ii}^{\rm T}=K$. Translated to $\mathcal{A}$, this means that $a_{ij}=0$ for all $i \neq j$ and the $a_{ii}$'s are identical orthogonal projection matrices of rank $Kp^{-1}$. This can be reached if and only if $\mathcal{J}_{i}=\mathcal{J}_{j}$ for all $i$ and $j$. It follows that 
\begin{equation}
\max_{\Omega} \left[ \left\{{\rm Tr}({\rm Tr}_{\rm R}P_{0}^{\vphantom{\rm T}})({\rm Tr}_{\rm R}P_{0}^{\rm T})\right\}_{\Sigma_{0}=\Omega} \right] \le pK,
\label{bound3c}
\end{equation}
where the equality sign is attained if and only if $\mathcal{J}_{i}=\mathcal{J}_{j}$ for all $i$ and $j$.

By combining the bounds of Eqs.\ (\ref{bound1c},\ref{bound2c},\ref{bound3c}), we obtain Eq.\ (\ref{ineqapp}). 

\section{Details about the simulation study}
\label{AppendixC}

In this Appendix, we give the algorithm used to calculate the maximum likelihood estimators. Also, additional simulation results are presented and $\delta$ is demonstrated to be sufficiently small.

\subsection{Calculating the maximum likelihood estimators}
\label{C1}

The CM algorithm is run with $\hat{\Sigma}_{1}=\openone_{p}$ and, after convergence is achieved ($|{\rm Det}\hat{\Sigma}_{n+1}-{\rm Det}\hat{\Sigma}_{n}| \le \delta  {\rm Det}\hat{\Sigma}_{n}$), we set $\hat{\Sigma}_{\rm temp}=\hat{\Sigma}_{n+1}$ and $\hat{B}_{\rm temp}=\hat{B}_{n}$. Then, another $\hat{\Sigma}_{1}$ is constructed by drawing a $p \times p$ matrix from $W_{p}(\openone_{p},p)$, where `$W$' denotes a Wishart distribution, and dividing it by $p$. With the randomly created $\hat{\Sigma}_{1}$, the CM algorithm is run up to convergence (possibly with another number of iterations than in the previous run) and if the newly calculated $\hat{\Sigma}_{\rm new}=\hat{\Sigma}_{n+1}$ and $\hat{B}_{\rm new}=\hat{B}_{n}$ satisfy 
\begin{equation}
\mathcal{L}(\hat{B}_{\rm new},\hat{\Sigma}_{\rm new}) > \mathcal{L}(\hat{B}_{\rm temp},\hat{\Sigma}_{\rm temp}) \quad \mbox{and} \quad |{\rm Det}\hat{\Sigma}_{\rm new}-{\rm Det}\hat{\Sigma}_{\rm temp}| > 10\delta {\rm Det}\hat{\Sigma}_{\rm temp},
\label{algineq}
\end{equation}
we set $\hat{\Sigma}_{\rm temp}=\hat{\Sigma}_{\rm new}$ and $\hat{B}_{\rm temp}=\hat{B}_{\rm new}$. The above is repeated until $\hat{\Sigma}_{\rm temp}$ and $\hat{B}_{\rm temp}$ remain unchanged for 10 different randomly created $\hat{\Sigma}_{1}$'s in a row. When the algorithm is terminated, $\mathcal{L}(\hat{B}_{\rm temp},\hat{\Sigma}_{\rm temp})$ is considered to be the global maximum of $\mathcal{L}(B,\Sigma)$ and we set $\hat{B}=\hat{B}_{\rm temp}$ and $\hat{\Sigma}=\hat{\Sigma}_{\rm temp}$. Table \ref{jumps} holds the number of jumps of $\hat{\Sigma}_{\rm temp}$ (and $\hat{B}_{\rm temp}$) in 1000 samples of size $N=15$ for $\rho=0.5$. (These are the same samples as the ones of Sec.\ \ref{simulation}.) It turns out that multi-modality is indeed rare: The candidate model with $i=j=5$ has a multi-modal $\mathcal{L}(B,\Sigma)$ in at most 1.6\% of the 1000 samples. Table \ref{jumps} also holds the number of additional $\hat{\Sigma}_{1}$'s in the 1000 samples. There are about 2 to 3 additional $\hat{\Sigma}_{1}$'s per jump and a maximum of 5 additional $\hat{\Sigma}_{1}$'s per jump. 

\begin{table}[h!]
\caption{Number of jumps of $\hat{\Sigma}_{\rm temp}$ and number of additional $\hat{\Sigma}_{1}$'s in 1000 samples of size $N=15$ for $\rho=0.5$.}  
\begin{tabular}{ccccccccccccc}  \hline 
   & & \multicolumn{5}{l}{jumps}              &   & \multicolumn{5}{l}{additional $\hat{\Sigma}_{1}$'s}    \\
i  & &  j   &        &         &       &      &   & j      &        &      &     &     \\ \cline{3-7} \cline{9-13}  
   & &  \,1\,   & \,2\,      & \,3\,       & \,4\,     & \,5\,    &   & \,1\,      &  \,2\,     & \,3\,    & 4   & 5   \\ \hline 
1  & &  0   & 0      & 0       & 0     & 0    &   & 0      &  0     & 0    & 0   & 0   \\ 
2  & &  0   & 0      & 0       & 1     & 1    &   & 0      &  0     & 0    & 5   & 2   \\  
3  & &  0   & 0      & 1       & 2     & 1    &   & 0      &  0     & 5    & 2   & 5   \\ 
4  & &  0   & 2      & 1       & 2     & 7    &   & 0      &  4     & 1    & 4   & 18  \\ 
5  & &  0   & 2      & 5       & 5     & 16   &   & 0      &  3     & 10   & 10  & 38  \\  \hline
\end{tabular}
\label{jumps}
\end{table} 

\subsection{Additional simulation results}
\label{C2}

The frequencies of selecting the correct model with AIC, AICc and BIC in 1000 samples of sizes $N=15,20,50$ are given in Table \ref{modelselection2} for $\rho = 0.2,0.5,0.8$. In the simulation, the values of the covariates in the samples of sizes $N=15$ and $N=20$ are the same as, respectively, the first 15 and 20 values of the covariates in the samples of size $N=50$. Table \ref{modelselection2} also holds the number of times that the difference between the second smallest and smallest value of a criterion is less than $200N\delta$ (10 times the numerical error of the difference). These numbers are of order unity such that $\delta$ is sufficiently small. The average (over 1000 samples) of the difference between the second smallest and smallest value of a criterion is not given in Table \ref{modelselection2}, but it ranges from 3.8 (for AIC with $N=15$ and $\rho=0.2$) to 52.4 (for BIC with $N=50$ and $\rho=0.8$). In all 1000 samples of sizes $N=20$ and $N=50$, there are no jumps of $\hat{\Sigma}_{\rm temp}$. In the samples of size $N=15$, the number of jumps of $\hat{\Sigma}_{\rm temp}$ and the number of additional $\hat{\Sigma}_{1}$'s do not depend much on $\rho$. 

\begin{table}[h!]
\caption{Frequencies $f$ and $\nu$ in 1000 samples of, respectively, selecting the correct model and the difference between the second smallest and smallest value of a criterion being smaller than $200N\delta$.}
\begin{tabular}{ccccccccc} \hline
N   &  $\rho$  &  \multicolumn{3}{c}{$f$} &  &  \multicolumn{3}{c}{$\nu$} \\ \cline{3-5} \cline{7-9}
    &          &  AIC  &  AICc  &  BIC    &  & AIC  &  AICc  &  BIC          \\ \hline
15  &  0.2     &  249  &  500   &  397    &  &  0   &  1     &  0            \\
    &  0.5     &  241  &  488   &  385    &  &  0   &  0     &  0            \\
    &  0.8     &  233  &  473   &  365    &  &  1   &  0     &  0            \\
20  &  0.2     &  366  &  570   &  577    &  &  0   &  0     &  0            \\
    &  0.5     &  353  &  604   &  611    &  &  0   &  2     &  0            \\ 
    &  0.8     &  324  &  553   &  556    &  &  0   &  0     &  0            \\
50  &  0.2     &  493  &  590   &  835    &  &  0   &  1     &  0            \\
    &  0.5     &  528  &  616   &  832    &  &  0   &  1     &  0            \\
    &  0.8     &  503  &  594   &  848    &  &  0   &  0     &  0            \\ \hline     
\end{tabular}
\label{modelselection2}
\end{table}


\begin{thebibliography}{99}

\bibitem{akaike73}
Akaike, H. (1973). Information theory and an extension of the maximum likelihood principle. In {\em 2nd International Symposium on Information Theory}, Ed. B. N. Petrov and F. Csaki, pp. 267-281. Budapest: Akademia Kiado.

\bibitem{hurvich89}
Hurvich, C. M. and Tsai, C. -L. (1989). Regression and time series model selection in small samples. {\em Biometrika} {\bf 76}, 297-307.

\bibitem{sugiura78} 
Sugiura, N. (1978). Further analysis of the data by Akaike's information criterion and the finite corrections. {\em Comm. Statist.} {\bf A7}, 13-26.

\bibitem{hurvich90}
Hurvich, C. M., Shumway, R. and Tsai, C. -L. (1990). Improved estimators of Kullback-Leibler information for autoregressive model selection in small samples. {\em Biometrika} {\bf 77}, 709-719.

\bibitem{hurvich93}
Hurvich, C. M. and Tsai, C. -L. (1993). A corrected Akaike information criterion for vector autoregressive model selection. {\em J. Time Ser. Anal.} {\bf 14}, 271-279.

\bibitem{bedrick94}
Bedrick, E. J. and Tsai, C. -L. (1994). Model selection for multivariate regression in small samples. {\em Biometrics} {\bf 50}, 226-231.

\bibitem{zellner62}
Zellner, A. (1962). An efficient method of estimating seemingly unrelated regressions and tests for aggregation bias. {\em J. Am. Statist. Assoc.} {\bf 57}, 348-368.

\bibitem{srivastava87}
Srivastava, V. K. and Giles, D. E. A. (1987). {\em Seemingly Unrelated Regression Equations Models}. New York: Marcel Dekker.

\bibitem{goldberger91}
Goldberger, A. S. (1991). {\em A Course in Econometrics}, p.\ 323. Cambridge, Massachusetts: Harvard University Press. 

\bibitem{verbyla88}
Verbyla, A. P. and Venables, W. N. (1988). An extension of the growth curve model. {\em Biometrika} {\bf 75}, 129-138.

\bibitem{rochon96}
Rochon, J. (1996). Analyzing bivariate repeated measures for discrete and continuous outcome variables. {\em Biometrics} {\bf 52}, 740-750.

\bibitem{andersson01}
Andersson, S. A., Madigan, D. and Perlman, M. D. (2001). Alternative Markov properties for chain graphs, {\em Scand. J. Statist.} {\bf 28}, 33-85. 

\bibitem{linhart86}
Linhart, H. and Zucchini, W. (1986). {\em Model Selection}, p.\ 245. New York: Wiley.

\bibitem{schwarz78}
Schwarz, G. (1978). Estimating the dimension of a model. {\em Ann. Statist.} {\bf 6}, 461-464.

\bibitem{meng93}
Meng, X.-L. and Rubin, D. B. (1993). Maximum likelihood estimation via the ECM algorithm: A general framework. {\em Biometrika} {\bf 80}, 267-278.

\bibitem{oberhofer74}
Oberhofer, W. and Kmenta, J. (1974). A general procedure for obtaining maximum likelihood estimates in generalized regression models. {\em Econometrica} {\bf 42}, 579-590.

\bibitem{drton04}
Drton, M. and Richardson, T. S. (2004). Multimodality of the likelihood in the bivariate seemingly unrelated regressions model. {\em Biometrika} {\bf 91}, 383-392.

\bibitem{drton06}
Drton, M. (2006). Computing all roots of the likelihood equations of seemingly unrelated regressions. {\em J. Symb. Comput.} {\bf 41}, 245-254.

\bibitem{hurvich91}
Hurvich, C. M. and Tsai, C. -L. (1991). Bias of the corrected AIC criterion for underfitted regression and time series models. {\em Biometrika} {\bf 78}, 499-509.

\bibitem{shibata80}
Shibata, R. (1980). Asymptotically efficient selection of the order of the model for estimating parameters of a linear process. {\em Ann. Statist.} {\bf 8}, 147-164.

\bibitem{shibata81}
Shibata, R. (1981). An optimal selection of regression variabales. {\em Biometrika} {\bf 68}, 45-54.


\end{thebibliography}
\end{document}